\documentclass[aps,prb,twocolumn,showpacs,superscriptaddress,preprintnumbers,amsmath,amssymb,longbibliography]{revtex4-1}
\usepackage{graphicx}
\usepackage{bm}
\usepackage{ulem}    
\usepackage{xcolor}
\usepackage{gensymb}                  
\usepackage{tikz}
\usepackage[pdftex,colorlinks=true,pdfstartview=FitV,linkcolor=blue,citecolor=blue,urlcolor=blue]{hyperref}

\begin{document}

\title{Simulation of thermodynamic properties of magnetic transition metals from \\
an efficient tight-binding model}
\author{Alexis Front}
\email{alexisfront@protonmail.com}
\affiliation{Laboratoire d'Etude des Microstructures, ONERA-CNRS, UMR104, Universit\'e Paris-Saclay, BP 72, Ch\^atillon Cedex, 92322, France}
\author{Georg Daniel F\"orster}
\email{georg-daniel.forster@univ-orleans.fr}
\affiliation{Laboratoire d'Etude des Microstructures, ONERA-CNRS, UMR104, Universit\'e Paris-Saclay, BP 72, Ch\^atillon Cedex, 92322, France}
\affiliation{Interfaces, Confinement, Mat\'eriaux et Nanostructures (ICMN), CNRS, Universit\'e d'Orl\'eans, Orl\'eans, France}
\author{Van-Truong Tran}
\affiliation{Universit\'e Paris-Saclay, CEA, Service de Recherches de M\'etallurgie Physique, 91191 Gif-sur-Yvette, France}
\author{Chu-Chun Fu} 
\affiliation{Universit\'e Paris-Saclay, CEA, Service de Recherches de M\'etallurgie Physique, 91191 Gif-sur-Yvette, France}
\author{Cyrille Barreteau}
\affiliation{DRF-Service de Physique de l'Etat Condens\'e, CEA-CNRS, Universit\'e Paris-Saclay, F-91191 Gif-sur-Yvette, France}
\author{Fran\c{c}ois Ducastelle}
\affiliation{Laboratoire d'Etude des Microstructures, ONERA-CNRS, UMR104, Universit\'e Paris-Saclay, BP 72, Ch\^atillon Cedex, 92322, France}
\author{Hakim Amara}
\email{hakim.amara@onera.fr}
\affiliation{Laboratoire d'Etude des Microstructures, ONERA-CNRS, UMR104, Universit\'e Paris-Saclay, BP 72, Ch\^atillon Cedex, 92322, France}
\affiliation{Universit\'e de Paris, Laboratoire Mat\'eriaux et Ph\'enom\`enes Quantiques (MPQ), CNRS-UMR7162, 75013 Paris, France}
%

\begin{abstract}

Atomic scale simulations at finite temperature are an ideal approach to study the thermodynamic properties of magnetic transition metals. However, the development of interatomic potentials explicitly taking into account magnetic variables is a delicate task.  In this context, we present a tight-binding model for magnetic transition metals in the Stoner approximation. This potential is integrated into a Monte Carlo structural relaxations code where trials of atomic displacements as well as fluctuations of local magnetic moments are performed to determine the thermodynamic equilibrium state of the considered systems. As an example, the Curie temperature of cobalt is investigated while showing the important role of atomic relaxations. Furthermore, our model is generalized to other transition metals highlighting a local magnetic moment distribution that varies with the gradual filling of the $d$ states. Consequently, the successful validation of the potential for different magnetic configurations indicates its great transferability makes it a good choice for atomistic simulations sampling a large configuration space. 

\end{abstract}

\maketitle 

\section{Introduction}

Magnetism plays a key role in many areas of materials science, especially when transition metals and their alloys are concerned. In these systems, magnetism can be the driving force impacting the phase stability and chemical ordering. Typical examples include the stability of the bcc $\alpha$ phase of iron,~\cite{Hasegawa1983, Herper1999} the phase diagram of Fe-Co~\cite{Abrikosov1996, Rahaman2011} or the Fe-Cr mixing enthalpy anomaly.~\cite{Paxton2008} However, the direct relationship, if any, between the atomic-scale origins of these properties and the contribution of magnetism remains a challenge nowadays.~\cite{Yesilleten1998, Mrovec2011} In this particular context, large-scale atomic simulations are required but they are still limited by the transferability of interatomic potentials including a magnetic contribution which is far from trivial.  A main difficulty lies in the establishment of a quantitative theory of finite temperature magnetism, which is still elusive and therefore represents an issue of both fundamental and applied importance. \\

In recent years, different kinds of interatomic potentials have been developed mainly to deal with the case of iron and its alloys, which represents a major issue in many steel industry applications.~\cite{Paxton2014} The majority of the existing interatomic potentials of Fe are based on the embedded atom method (EAM) or the Finnis-Sinclair model with a more or less precise description of the directional bonds.~\cite{Dudarev2005, Muller2007, Finnis1984, Ackland2004, Mishin2005, Hepburn2008, Starikov2020} Meanwhile, a better treatment of magnetism can be obtained by coupling classical empirical potentials and Heisenberg-type models for spin dynamics.~\cite{Ma2017, Tranchida2018} Aside from the classical potentials is the tight-binding (TB) framework that allows an explicit dealing with the electrons making magnetism a natural consequence of the model. In addition, they have the advantage of being transparent and simple, while still allowing for a high degree of transferability to handle magnetic systems.~\cite{Barreteau2016, Soin2011, Drautz2011, Ford2014, Goyhenex2016} Despite their success, these different types of interatomic potentials (i.e. empirical or TB)  have so far been applied mostly to study the stability of bulk phases at 0 K and also to deal with some specific defects (point defects,~\cite{Liu2005, Paxton2010, Ford2014} dislocations,~\cite{Mrovec2011} grain boundaries,~\cite{Yesilleten1998, Starikov2020} ...) which are crucial for the use of magnetic materials in various applications. However, the case of thermodynamic properties at finite temperature, which is much more complex, is still elusive.~\cite{Starikov2021} A challenge for such simulations is to have an energy model able to describe magnetic phase transformations where atomic relaxations are included to study relatively large systems. 

More precisely, it is difficult to develop a model for transition metals describing a local electron-electron interaction that is strong enough to create a localized magnetism fluctuating on a short length scale or to give rise to significant hybridization of the $d$ states with the surrounding atoms resulting in an itinerant magnetism. In practice, these different magnetic degrees of freedom can be accurately described either by a localized Heisenberg model (rare earths or transition metal insulators) or in the framework of the pure $d$ band Stoner theory. Consequently, the development of a unifying model remains scarce.~\cite{Moriya1978, Moriya1985, Ruban1} In this context, a typical challenge is to develop a theory able to produce accurate Curie temperatures (T$_{\mathrm{C}}$) for $d$ elements. Back in the 60's, Friedel \textit{et al.} had already proposed a simple model of magnetism for transition metals which is somewhat intermediate between the Heisenberg atomic model and the Stoner band model.~\cite{Friedel1961} Further theoretical works have been developed within an \textit{ab initio} framework based on the disordered local moment (DLM) approximation as an accurate representation of a paramagnetic configuration with a random alloy of spin-up and spin-down atoms.~\cite{Gyorffy1985} As an example, the DLM is used to get parameters of a magnetic model and the temperature dependence properties are determined by Monte Carlo. Within this approach, the calculated Curie temperatures and paramagnetic susceptibilities were found in good agreement with experimental data for bcc Fe and fcc Ni.~\cite{Ruban1} Recently, an \textit{ab-initio}-based effective interaction model (EIM) has been developed for the study of magnetism, chemical-phase stability and their coupling in bcc Fe-Co structures.~\cite{Tranb, Schneider2020} The EIMs can be quite efficient to treat thermodynamic and kinetic properties, but the lattice-vibration effects are not considered explicitly.~\cite{Schneider2021} Despite such intense efforts,~\cite{Kormann2016} these approaches are not adapted to the development of interatomic potentials to investigate the structural properties of magnetic transition metals at finite temperature where large systems and complete relaxation of the system are required. \\
 
In this work, we present a tight-binding interatomic potential including all relevant physics related to collinear magnetism in transition metals while remaining simple enough to allow the simulation of hundreds or even thousands of atoms at finite temperature. A fourth-moment approximation to the local density of states developed for transition-metal carbides~\cite{Amara2009, Los2011, AguiarHualde2017} is extended to take into account explicit magnetic contribution via the Stoner theory of itinerant magnetism.~\cite{Stoner1939} This semi-empirical approach relies on local (atomic) energy calculations using the recursion method and is coupled with Monte Carlo (MC) simulations in order to relax the structures and calculate thermodynamic properties. The paper is organized as follows. In Sec.~\ref{Section_TB}, we present the tight-binding approximation coupled to the Stoner model developed to calculate band energies including a magnetic contribution. Empirical repulsive terms are then added to obtain total energies. The Monte Carlo procedure used to relax the structures is also described. Different validations and applications of the model to determine the Curie temperature of Co are developed and discussed in Sec.~\ref{Section_Co}. Lastly, Sec.~\ref{Sec_Generalized_model} is devoted to the generalization of our tight-binding model to other magnetic transition metals. 

\section{Tight-binding Hamiltonian including the Stoner model}
\label{Section_TB}

\subsection{Tight-binding Hamiltonian with various approximations}

There are several magnetic TB models to characterize transition metals and their alloys that contrast with the Hamiltonian approximation level. This mainly concerns the choice of the basis which may be orthogonal~\cite{McEniry2011} or not~\cite{Paxton2008} and includes different orbitals ($spd$~\cite{Barreteau2016} or only $d$~\cite{Ducastelle1970, Drautz2006} for pure transition metals). As it is a parameterized quantum description, the evaluation of complex integrals is avoided and replaced by functionals whose form and parameterization differ according to the TB model. 
In case of interatomic potentials, the total energy (with respect to the energy of the free atoms) can be written as the sum of an attractive contribution, which describes the formation of an energy band when atoms are put together to form a density of states, and of a phenomenological repulsive term, which empirically accounts for the ionic and electronic repulsions.~\cite{Ducastelle1991, Pettifor1995} The magnetic term is introduced via the Stoner Hamiltonian to remove the degeneracy between the two spin directions through a potential that generates a band splitting between "up" and "down" spins.~\cite{Stoner1939} At this stage, there are different approaches to develop interatomic potentials from the TB Hamiltonian. The most specific and standard one is basically to perform a complete diagonalization of the Hamiltonian to get an accurate contribution of the band term. However, the price to pay is usually quite high in terms of computational time, especially if the model is implemented in Monte Carlo or Molecular Dynamic code to relax structures where many steps are required to converge. To overcome this difficulty, it is possible to obtain a simplified model of the band term using the moment method~\cite{Gaspard1973} or recursion method.~\cite{Haydock1972, Pettifor1985} A decomposition of band energy into binding energies can be derived which, combined with the theory of perturbations with respect to the underlying electronic structure, results in analytical bond-order potentials (BOPs).~\cite{Pettifor1995, Drautz2006, Hammerschmidt2019} Being limited to numerical calculations of the first moments of the local density of states, the calculation becomes fast and can be integrated into structural relaxation codes.~\cite{Amara2009} Within this framework, Ackland \textit{et al.}~\cite{Ackland2006} have established a magnetic interatomic potential where the band term was calculated within the second-moment approximation of the TB model. Unfortunately, the limited description of the density of states does not allow to account for the subtle relationship that can exist between magnetism and the structural stability of pure transition metals and their alloys.~\cite{Karoui2013} \\

In the following, a magnetic interatomic potential based on the tight-binding framework is detailed which provides an efficient way to calculate the structural properties of magnetic transition metals at finite temperature. The model based on the fourth-moment approximation for the band term is the simplest in terms of moments that still correctly describes some of the magnetic features.

\subsection{Fourth-moment approximation}

There is no need to detail the tight-binding approximation here. The technical and theoretical aspects concerning our model in the fourth-moment approximation (FMA) to handle transition-metal carbides as well as its transferability are given in the Ref.~\citenum{Amara2009}. In the following section are summarized the essential points that are relevant to understand the extension of our model to take into account explicit magnetic contribution via the Stoner theory. \\

In case of a non-magnetic (NM) system containing $N$ atoms, the total energy of an atom $i$ ($E_{\mathrm{tot/NM}}^{i}$) is divided into two contributions, a band structure term ($E_{\mathrm{band}}^{i}$) that describes the formation of an energy band when atoms are put together and a repulsive term ($E_{\mathrm{rep}}^{i}$) that empirically accounts for ionic and electronic repulsions:
\begin{equation}
E_{\mathrm{tot/NM}}^{i}= E_{\mathrm{band}}^{i} + E_{\mathrm{rep}}^{i}
\label{tot_NM_i}
\end{equation}
The total energy of the system, $E_{\mathrm{tot/NM}}$, then writes:
\begin{equation}
E_{\mathrm{tot/NM}} = \sum_{i\ \mathrm{atoms}}E_{\mathrm{tot/NM}}^{i}
\label{tot_NM}
\end{equation}
The band term is given by the following equation: 
\begin{equation}
E_{\mathrm{band}}^{i} =\int^{\epsilon_{\mathrm{F}}}_{-\infty}(E-\epsilon_{i}^{0})n_{i}(E)dE
\label{band}
\end{equation}
where $\epsilon_{\mathrm{F}}$ is the Fermi level and $\epsilon_{i}^{0}$ the atomic energy level. We use the recursion method to calculate $n_{i}(E)$, the local electronic density of states (LDOS) on each site $i$.~\cite{Haydock1972} Only the first four continued fraction coefficients are calculated exactly, which provides already a good description of the angular contributions to the energy and leads to a relatively fast scheme of order $N$. Since we want to have a consistent and simple scheme to describe correctly the transition elements, the fourth moment approximation is a good compromise. Taking into account the fifth or sixth moment would be even better but rather expensive in terms of computational effort.~\cite{Turchi1985, Nastar1995} To keep the model as simple and fast as possible, we neglect the $sp$ electrons that form a broader nearly-free-electron band. Only the $d$ electrons are taken into account as long as we are interested in cohesive properties more than in a detailed description of the electronic structure.  Indeed, the bell-shape behavior of the cohesive energy and of the elastic moduli is correctly predicted by the TB approximation where $sp-d$ hybridizations are neglected and is the consequence of a gradual filling of the $d$ states.~\cite{Ducastelle1970} When interested in more detailed electronic structure properties, $sp-d$ hybridization should however be taken into account as done in Ref.\citenum{Barreteau1998}. Thus, we will work with the $|i\mu \rangle$ basis where $\mu$ is the orbital index ($\mu = d_{xy}, d_{yz}, d_{zx}, d_{x^{2}-y^{2}}, d_{3z^{2}-r^{2}}$). In our $d$ band model, the Slater-Koster parameters for the hopping integrals $dd\sigma, dd\pi$, and $dd\delta$ are considered to be in the ratio -2:1:0 and to decay exponentially with respect to the distance $r$ as: 
\begin{equation}
dd\lambda(r)=dd\lambda_{0}\exp\left[-q\left(\frac{r}{r_{0}}-1\right)\right] \quad,
\label{hopping_integrals}
\end{equation}
where $\lambda = \sigma, \pi, \delta$. In case of metallic systems, it is common to impose a condition of local charge neutrality which can be achieved by locally varying atomic energy levels. Instead of following this procedure, a more approximate but much easier scheme used here  is to calculate the total energies based on a local charge neutrality hypothesis by introducing fictious local Fermi levels. The second term in Eq. \ref{tot_NM_i}, $E_{\mathrm{rep}}^{i}$, is a repulsive contribution chosen to have a pairwise Born-Mayer form here:
\begin{equation}
E^{i}_{\mathrm{rep}}=A\sum_{j\neq i}\exp\left[-p\left(\frac{r_{ij}}{r_{0}}-1\right)\right]
\label{repulsif}
\end{equation}
In our TB model based on a fourth-moment approximation, magnetism is introduced via the Stoner model~\cite{Stoner1939} by including the presence of local exchange fields within the band energy of Eq.~\ref{band}. We place ourselves in the case of collinear magnetism that imposes the differentiation of two spin populations: spin up ($\uparrow$) and spin down ($\downarrow$). The
spin moment $m_{i}$ in $\mu_{B}$ units is given by:
\begin{equation}
m_{i} = N_{i}\uparrow-N_{i}\downarrow
\label{momentmagnetic}
\end{equation}
where $N_{i}\uparrow$ and $N_{i}\downarrow$  are respectively the number of electrons in majority and minority spin bands of an atom $i$. The exchange potential is modeled by an effective magnetic field of the form: $Im_{i}/2$ where $I$ is the Stoner exchange integral. This allows us to define local magnetic on-site levels: 
\begin{equation}
\epsilon_{i}^{\sigma} = \epsilon_{i}^{0} \pm \frac{I}{2}m_{i}
\label{atomiclevelshift}
\end{equation}
The minus (plus) sign is chosen if the spin $\sigma$ is parallel (antiparallel) to the direction of the local magnetic field. From Eq.~\ref{atomiclevelshift}, it is obvious that these levels must be determined self-consistently, since the TB Hamiltonian now depends on the local magnetic moments. A straightforward procedure is to start from an initial guess for $m_{i}$, and once the corresponding Hamiltonian is diagonalized, two density of states (for up and down states) are obtained leading to an improved estimation. This latter is used as new input and the process is iterated until convergence. In practice, self-consistent magnetic moments were found using the Broyden mixing scheme.~\cite{Broyden1965} After summing over the whole electron population with the consideration of the double counting of states, the result is a contribution of this exchange potential to the total energy of an atom $i$:
\begin{eqnarray}
\label{Etotmag}
E_{\mathrm{tot}}^{i} &=& E_{\mathrm{band}}^{i} + E_{\mathrm{rep}}^{i} + E_{\mathrm{exc}}^{i} \\
E_{\mathrm{exc}}^{i} &=& -\frac{I}{4}m_{i}^{2}
\end{eqnarray}
where $E_{\mathrm{exc}}^{i}$ is the exchange energy.~\cite{Pettifor1995, Liu2005, Ford2014} 

\subsection{Monte Carlo simulations}

\begin{figure}[htbp!]
\begin{center}
\includegraphics[width=1.0\linewidth]{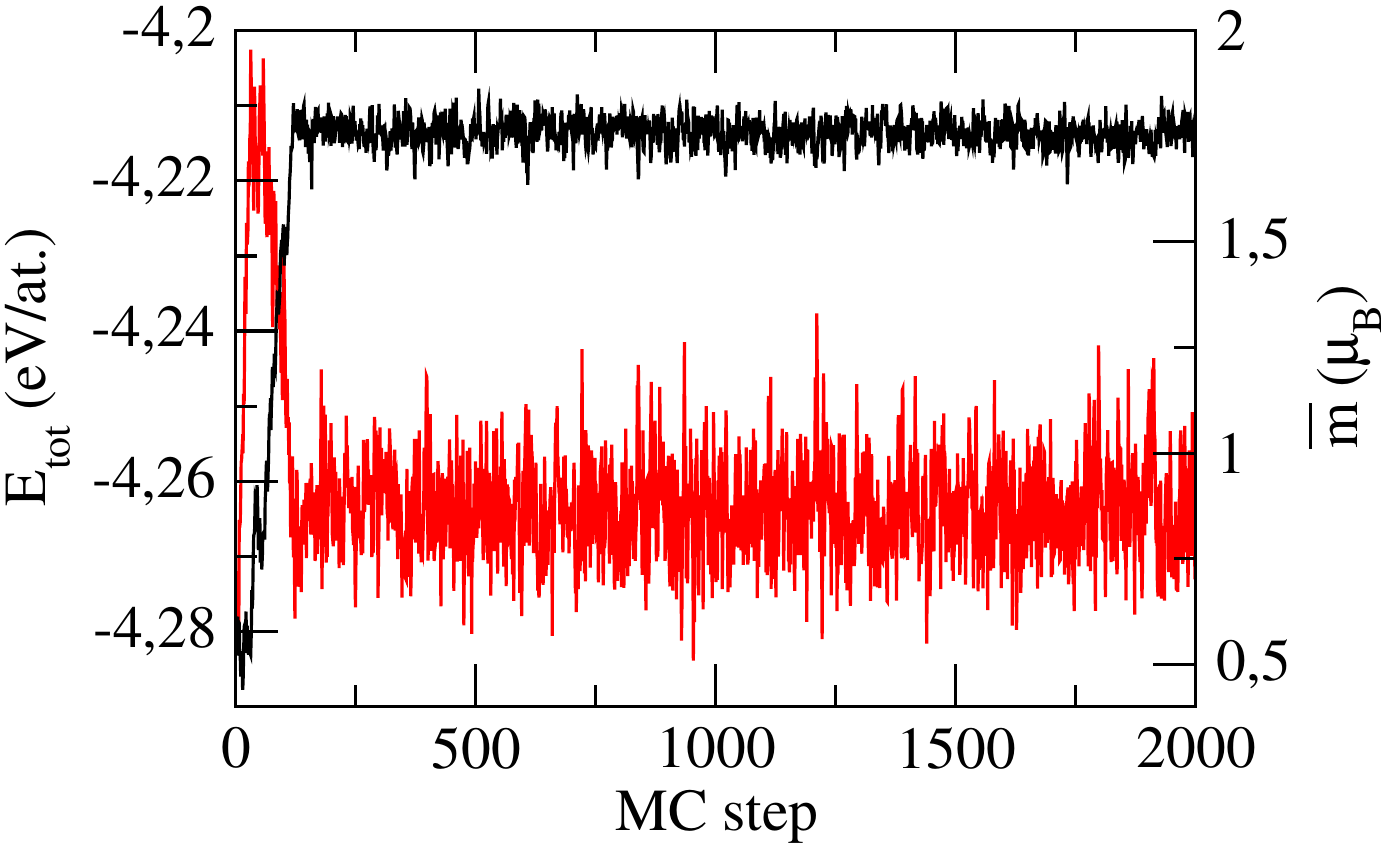}
\caption{Total energy (in red) and magnetic moment (in black) as a function of the number of MC steps at 800 K starting from a fcc random spin configuration.}
\label{fgr:e_m_conv}
\end{center}
\end{figure}

This atomic interaction model is then implemented in a Monte Carlo (MC) code, based on the Metropolis algorithm,~\cite{Metropolis1953} using the canonical ensemble.~\cite{Frenkel2002} This procedure makes it possible to relax the structures at finite temperatures according to a Boltzmann type probability distribution. In the canonical ensemble, standard MC trials correspond to random displacements. A MC macrostep corresponds to $N$ propositions of random atomic displacements, $N$ being the total number of atoms of the system. In principle, the determination of all local magnetic moments for each trial configuration is based on a brute force method which consists in performing two self-consistent calculations to extract at the end a very small energy difference. However, at finite temperature we allow for fluctuations of the magnetic moment. This renders the self-consistent determination of magnetic moments not suitable. To tackle this difficulty, a MC trial corresponds to randomly choosing an atom and its displacement as well as its local magnetic moment with an amplitude of $0.05\sqrt{T}(\xi-0.5)$ and $0.4\ln{T}(\xi-0.5)$ respectively. $T$ is the temperature in eV units and $\xi$ is a random number between 0 and 1. Both magnitudes have been adjusted in order to have 50\% of the MC moves accepted. For each run, we check the convergence of the total energy and the average magnetic moment, defined as $\bar{m}=1/N\sum_{i}^{N} m_{i}$ (see Fig.~\ref{fgr:e_m_conv}). Starting from a fully random magnetic state, the system converges rapidly to a ferromagnetic state corresponding to $\bar{m}=1.87$ $\mu_{B}$. We may notice that depending on the MC moves, the symmetrical value (-1.87 $\mu_{B}$) is also possible since these two magnetic states are degenerate. We performed $10^{3}$ MC macrosteps for equilibration then the average quantities are calculated over $10^{3}$ macrosteps. Since the total energy is taken as a sum of local terms, this avoids recalculating the total energy and the magnetic contribution of the whole system at each step of the Monte Carlo process making efficient the implementation of our TB model. Consequently, the local energy is only recalculated at each MC trial for atoms impacted by the displacement of an atom $i$. This approach is then perfectly adapted to deal with large systems and to reproduce the main energy properties of magnetic transition metals. 
\subsection{Fitting procedure}

The problem of finding a good parameter set for a TB interatomic model corresponds to an optimization problem, where one tries to reproduce a database by adjusting the model parameters: $dd\sigma$, $q$, $r_0$, $A$, $p$, $I$, the number of electrons $N_{d}$ and an inner $r_c^i$ and outer cut-off radius $r_c^o$. The latter are involved in the cut-off function that is applied to the hopping integrals and the repulsive energy:

\[
f_{\rm cut}\left(r\right) =
\begin{cases}
1                                                                         & {\rm if} \quad r \leq r_c^i \\ 
\frac{1}{2}\left[1+\cos\left(\pi\frac{r-r_c^i}{r_c^o-r_c^i}\right)\right] & {\rm if} \quad r_c^i < r < r_c^o \\
0                                                                         & {\rm if} \quad r \geq r_c^o
\end{cases}
\]

With a TB model, the parameters have a physical meaning limiting the ranges over which they can be optimized, reducing significantly the search space. The reference data are obtained by performing \textit{ab initio} calculations (Vienna Ab-initio Simulation Package (VASP)~\cite{Kresse1999}) using density functional theory (DFT) (details in Ref.~\citenum{Tran2020}), and contain the cohesive energies, magnetic moments, lattice constants, and elastic constants of non-magnetic and magnetic calculations for fcc and bcc crystals. Instead of evaluating explicitly these observables, we follow the data points on the strain-energy and strain-magnetization curves as computed by DFT. This approach allows for fewer evaluations of the potential energy leading to an exact comparison with the DFT results. The objective function that we optimize is the root mean square difference between the DFT results and the data obtained with a particular trial parameter set for the TB model. With 9 parameters the parameter space is still too large for a systematic exploration. For instance, sampling all combinations, while trying 10 values per parameter, $10^9$ evaluations of the reference data would be required. We therefore resort to parallel tempering Monte Carlo simulations~\cite{Earl2005} for this global optimisation problem. Making efficient use of parallel computers, the idea of this approach is that the high-temperature thermostats explore large regions of the parameter space, while those at lower temperature explore in detail local optima. The exchanges allow to escape bad local optima when better ones are discovered at a higher temperature. We use optimization runs with 10$^{5}$ MC cycles with 12 thermostats each with a 50\% temperature difference with the neighboring thermostat. Exchanges are attempted every 50 steps. The values of the maximally attempted change of the parameters is adjusted in each thermostat during the optimization such that about half of the MC moves are accepted. The best parameter set obtained by these runs was further locally optimized by the Nelder-Mead algorithm~\cite{Nelder1965} and scrutinized manually. 

\section{Structural properties of Co}
\label{Section_Co}

\subsection{Ground state at 0 K}

Experimentally, the hexagonal compact (hcp) structure is the most stable configuration for Co at low temperature with a hcp-fcc transition occuring at 680 K.~\cite{Hultgren} Since we want to investigate the Curie temperature ($\sim$ 1400 K), we focus our investigation on the fcc phase as well as the bcc one for comparison. Following our fitting procedure, the final parameter set corresponding to both structures for non-magnetic (NM) and ferromagnetic (FM) states can be found in Tab.~\ref{tab:parameters}. 
\begin{table}[htbp!]
\begin{center}
\setlength{\tabcolsep}{2pt}
\begin{tabular}{c|c|c|c|c|c|c|c|c}
$dd\sigma$&$q$&$r_{0}$&$A$&$p$&$I$&$N_{d}$&$r_{c}^{i}$&$r_{c}^{o}$ \\
\hline
\hline
1.39 &2.50 & 2.21 & 0.246 & 12.4 & 1.38 & 8.13 & 2.6 & 3.4 \\
\end{tabular}
\end{center}
\caption{Co parameters for the magnetic TB-FMA model, obtained by fitting to DFT reference data. $dd{\sigma}$, $A$ and $I$ are in eV, $r_{0}$, $r_{c}^{i}$ and $r_{c}^{o}$ are in~\AA.}
\label{tab:parameters}
\end{table}
\begin{figure}[htbp!]
\begin{center}
\includegraphics[width=0.90\linewidth]{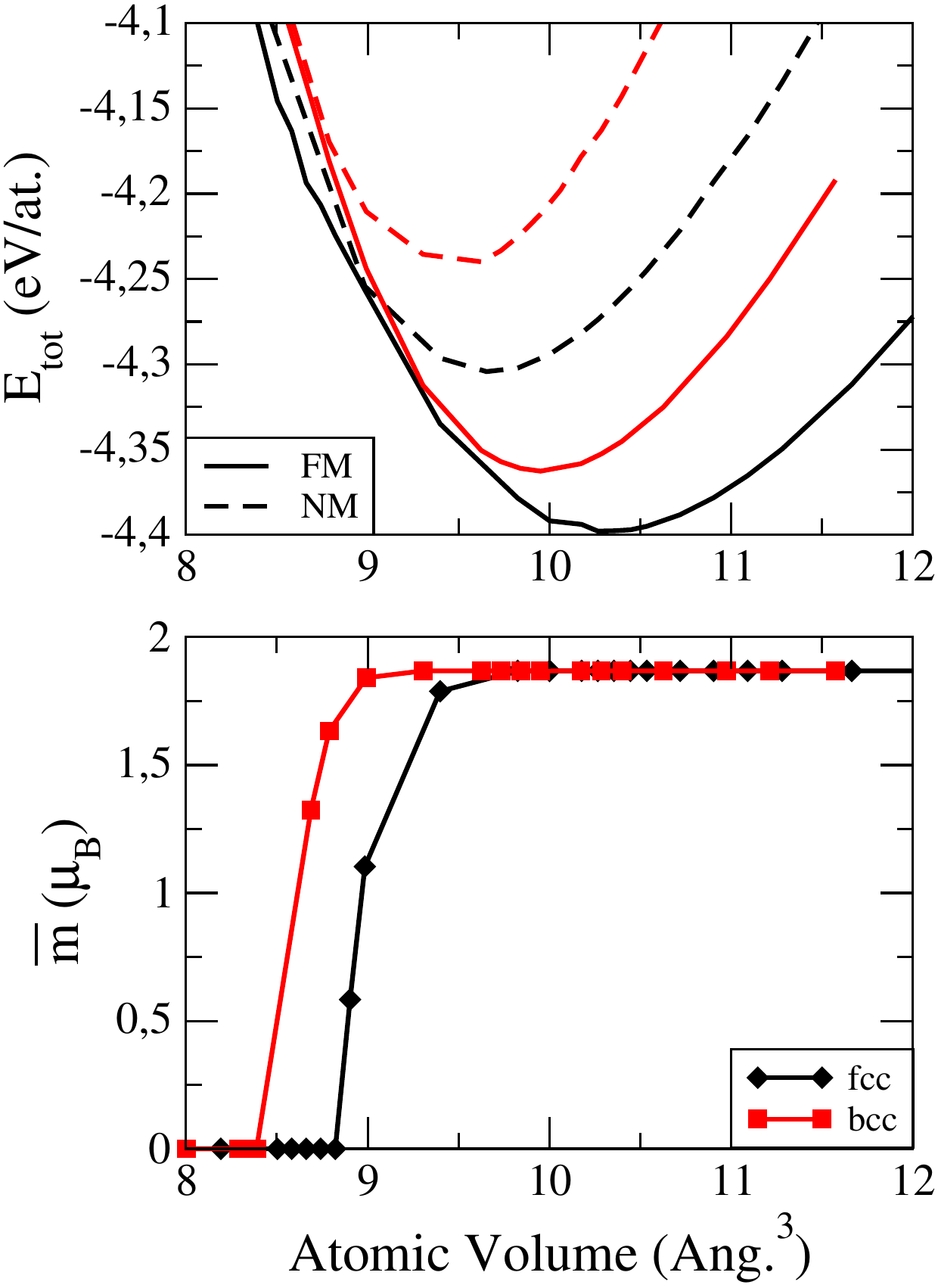}
\caption{Total energy (top) and total magnetic moment (bottom) as a function of atomic volume for fcc (in black) and bcc (in red) at 0 K calculated by TB-FMA. On the top, dashed lines represent a non-magnetic system and full lines a ferromagnetic system.}
\label{fgr:0K_E_m_V}
\end{center}
\end{figure}
The relative stability of these various phases, as well as the influence of magnetism on the system can be determined from the energy versus atomic volume curves plotted in Fig.~\ref{fgr:0K_E_m_V}. For calculations at 0 K, $m$ is computed self-consistently while the MC procedure defined previously will be privileged for the simulations at finite temperature. As illustrated in Table~\ref{tab:0K_E_I}, our model predicts a fcc ground state for both non-magnetic and ferromagnetic states in agreement with DFT calculations. More precisely, our TB model tends to stabilize the magnetic structure as in DFT with lattice parameters and cohesive energies always larger than in the non-magnetic phase.~\cite{Karoui2013} It is worth mentioning that our model reproduces well experimental elastic constants of the fcc structure. Regarding the magnetic moment, as expected, it increases when the lattice is expanded and vanishes when it is reduced.~\cite{Tran2020,Cerny} Meanwhile, the bcc magnetic moment appears for smaller atomic volume. Moreover, the value of the magnetic moment equal to 1.87 $\mu_{B}$ is completely determined in our TB formalism by $N_{d}$ the total number of electrons in our model ($m=10-N_{d}$).~\cite{Mrovec2011}
\begin{table*}[htbp!]
\begin{center}
\setlength{\tabcolsep}{2pt}
\begin{tabular}{c|c|c|c||c|c|c}
\multicolumn{2}{c|}{}   & NM & NM &FM &FM &FM \\
\multicolumn{2}{c|}{}   & DFT & TB-FMA &exp. &DFT & TB-FMA \\
\hline    BCC & $a$ (\AA)  & 2.76 & 2.68 & - & 2.81 & 2.71 \\
    & $E_{\mathrm{coh}}$ (eV/at.)  & -3.97 & -4.24 & - & -4.30 & -4.36 \\
\hline   FCC & $a$ (\AA)  & 3.45 & 3.38 & 3.54 \cite{Andreazza} & 3.52 & 3.45 \\
   & $E_{\mathrm{coh}}$ (eV/at.)  & -4.20 & -4.30 & -4.39\cite{Andreazza} (hcp) & -4.40 & -4.40 \\
    & $C_{11}$, $C_{12}$, $C_{44}$ (GPa)  & 407, 180, 208 & 389, 235, 142 & 225, 160, 92  \cite{Gump} & 290, 170, 145 & 273, 175, 109 \\
  \hline
\end{tabular}
\end{center}
\caption{DFT and TB-FMA calculations of the lattice parameters, cohesive energies and elastic constants for non-magnetic and ferromagnetic bcc and fcc systems at 0 K. Experimental data are only available for fcc FM phases. DFT calculations of bcc and fcc systems are extracted from Ref.\citenum{Tran2020}. }
\label{tab:0K_E_I}
\end{table*}

Besides the cohesive properties, the exchange interaction $J$ is a good descriptor to have an idea of the Curie temperature trend in the different structures. Indeed, $J$ is defined as the energy difference between a configuration with all spins up (or down) and a configuration where one spin is flipped. The value of $J$ is larger for the bcc structure (286 meV) than for the fcc one (178 meV) suggesting that the bcc system should have a $\mathrm{T_{C}}$ larger than the fcc one. \\

Once all parameters have been fitted and validated at 0 K, the difficulty in the derivation of a complete interatomic potential is to confirm its robustness at finite temperature to check its transferability. In the following, we go further by studying the Curie temperature of fcc and bcc systems performing off-lattice MC simulations where all degrees of freedom are considered. More precisely, each physical ingredient will be integrated step-by-step to determine its impact on the $\mathrm{T_{C}}$ calculation, i.e. longitudinal spin fluctuations, lattice vibrations and lattice expansion.

\subsection{Finite temperature}

\begin{figure}[htbp!]
\begin{center}
\includegraphics[width=0.95\linewidth]{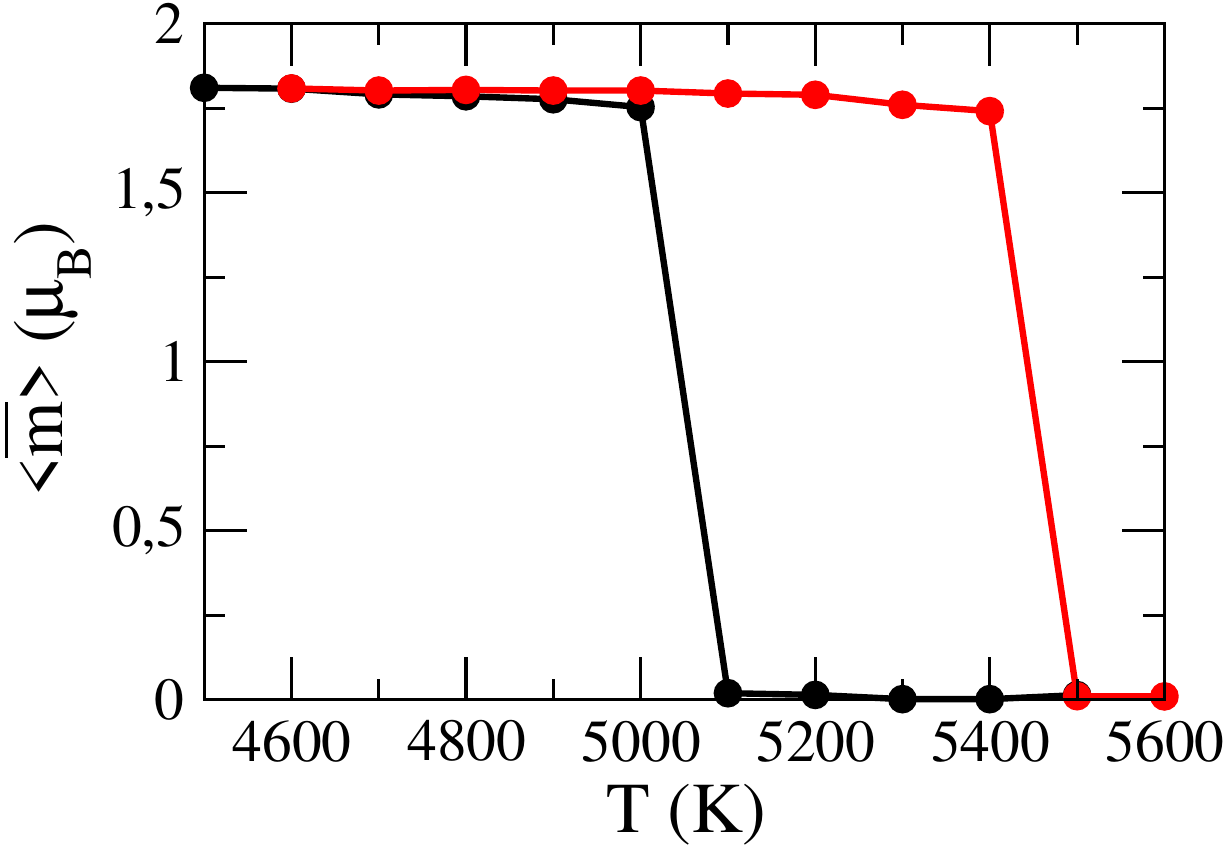}
\caption{Curie temperature of fcc (in black) and bcc systems (in red): total magnetic moment average as a function of temperature on Ising-type lattice.}
\label{fgr:Ising}
\end{center}
\end{figure}

\begin{figure}[htbp!]
\begin{center}
\includegraphics[width=0.95\linewidth]{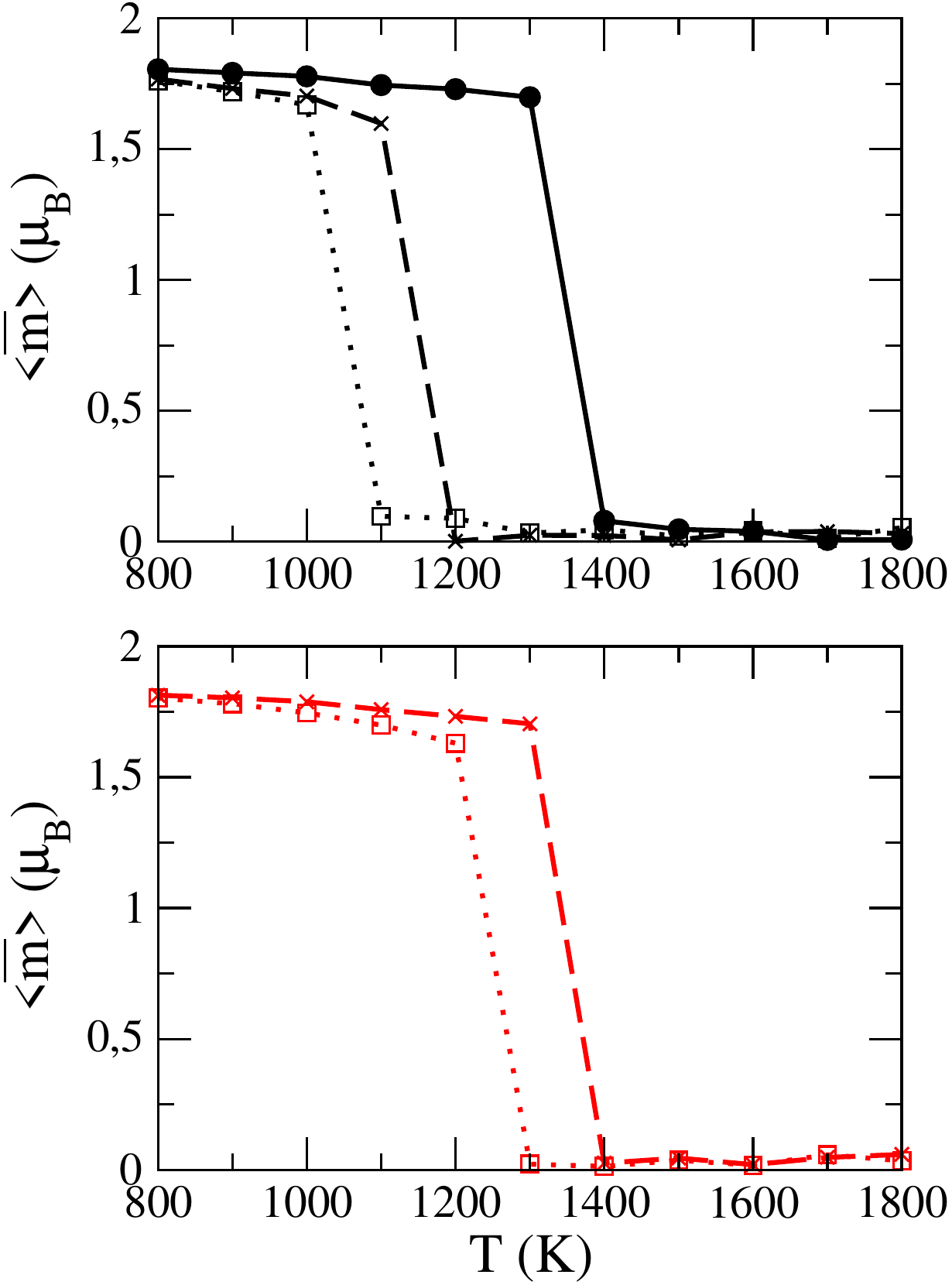}
\caption{Curie temperature of fcc (top) and bcc system (bottom): total magnetic moment average as a function of temperature with different approximations: longitudinal spin fluctuations (dashed line with cross), lattice vibrations (dotted line with square) and lattice expansion (full line with circle). }
\label{fgr:Tc}
\end{center}
\end{figure}

First, the Curie temperature is investigated on a rigid lattice where the MC trial consists in flipping the magnetic moment ($m=\pm1.87\text{\space}\mu_{B}$). In these Ising-type simulations, we considered a fcc/bcc system of 256/250 atoms which is sufficiently large for the convergence of the energy and the magnetization. We performed heating/cooling (increasing/decreasing temperatures) MC simulations which means the simulation at the next temperature starts from the last converged configuration of the previous temperature. No difference between the increasing and decreasing temperatures was observed, so that only the increasing temperature is shown in Fig.~\ref{fgr:Ising}. As expected, the Curie temperature for the bcc phase is larger than for the fcc one. However, the calculated $\mathrm{T_{C}}$ are really large compared to the experimental values ($\mathrm{T_{C}}=1388$ K for fcc~\cite{Tranb}) This is not surprising because magnetic moment fluctuations, lattice vibrations and lattice expansion are not considered. 
We now investigate the impact of magnetic moment fluctuations on the Curie temperature. As before, a fcc/bcc system containing 256/250 atoms on a rigid lattice is considered. In contrast to the previous case, the local magnetic moment is no longer constrained to two values but is free to fluctuate randomly in a continuous manner. The equilibrium is consequently longer to reach since both total energy and magnetic moment require more MC steps to converge. The results for the fcc and bcc phases are presented in Fig.~\ref{fgr:Tc}. Compared to the Ising-type simulation, we can clearly notice a spectacular effect of the magnetic fluctuations since $\mathrm{T_{C}}$ is drastically reduced by about 3500 K. Indeed, the magnetic moment decreases slowly before the Curie temperature equal to 1150 K and 1350 K for fcc and bcc systems respectively, followed by an abrupt drop to zero typical of a first order transition. It can be noted that a second order transition is observed within an Ising model where large boxes of simulations are considered (several thousand atoms). This is not the case in the present work which is focused to boxes containing several hundred atoms and therefore cannot reproduce a second order phase transition.~\cite{Diep1989}. Nevertheless, these deviations do not prevent us from investigations physical properties of magnetic transition metals at finite temperatures. Improving the accuracy would imply increasing the size of the systems which would make the calculations very time consuming and would be not adapted for highlighting the generalities of our model.

Interestingly, the impact of magnetic moment fluctuations is therefore crucial but insufficient to reproduce the experimental $\mathrm{T_{C}}$ value ($\sim$1400 K for the fcc phase). To go beyond, lattice vibrations (atomic relaxations) as well as magnetic fluctuations are taken into account to highlight the role of phonons and magnon-phonon coupling on the $\mathrm{T_{C}}$. As seen in Fig.~\ref{fgr:Tc}, the Curie temperature is slightly reduced ($\sim$ 100 K) to reach 1050 K and 1250 K for fcc and bcc systems, respectively. It may seem surprising or even disappointing that improving the model by adding the magnon-phonon coupling further deteriorates the prediction of the Curie temperature. However, this tendency is in good agreement with a recent first-principles thermodynamic approach developed for investigating the effect of phonons on magnetism for bcc Fe.~\cite{Tanaka2020} In this elegant work, the authors point out that the phonon softening due to magnetic disordering leads to the stabilisation of paramagnetic states resulting in a decrease of $\mathrm{T_{C}}$ by nearly 560 K.
Although this so-called feedback effect is not as strong in our case (here cobalt), it tends to prove the robustness of our model which in a natural and simple way is able to reproduce the rather complex physics discussed in Ref.\citenum{Tanaka2020}. Lastly, off-lattice MC simulations are performed including lattice expansion as well as magnetic moment fluctuations and atomic relaxations. As seen in Fig.~\ref{fgr:Tc}, no results are reported for the bcc phase. During the simulation, the lack of constraints on the simulation box makes possible the phase transformation to the most stable structure, the fcc one. Consequently, this result shows the efficiency of our interatomic potential to characterize the thermodynamic properties of magnetic systems. Regarding the fcc phase, the lattice contribution to $\mathrm{T_{C}}$ is significant with an increase of about 300 K. As a result, our TB model predicts a Curie temperature around 1350 K which is in good agreement with the experimental value.  \\
\begin{figure}[htbp!]
\begin{center}
\includegraphics[width=0.95\linewidth]{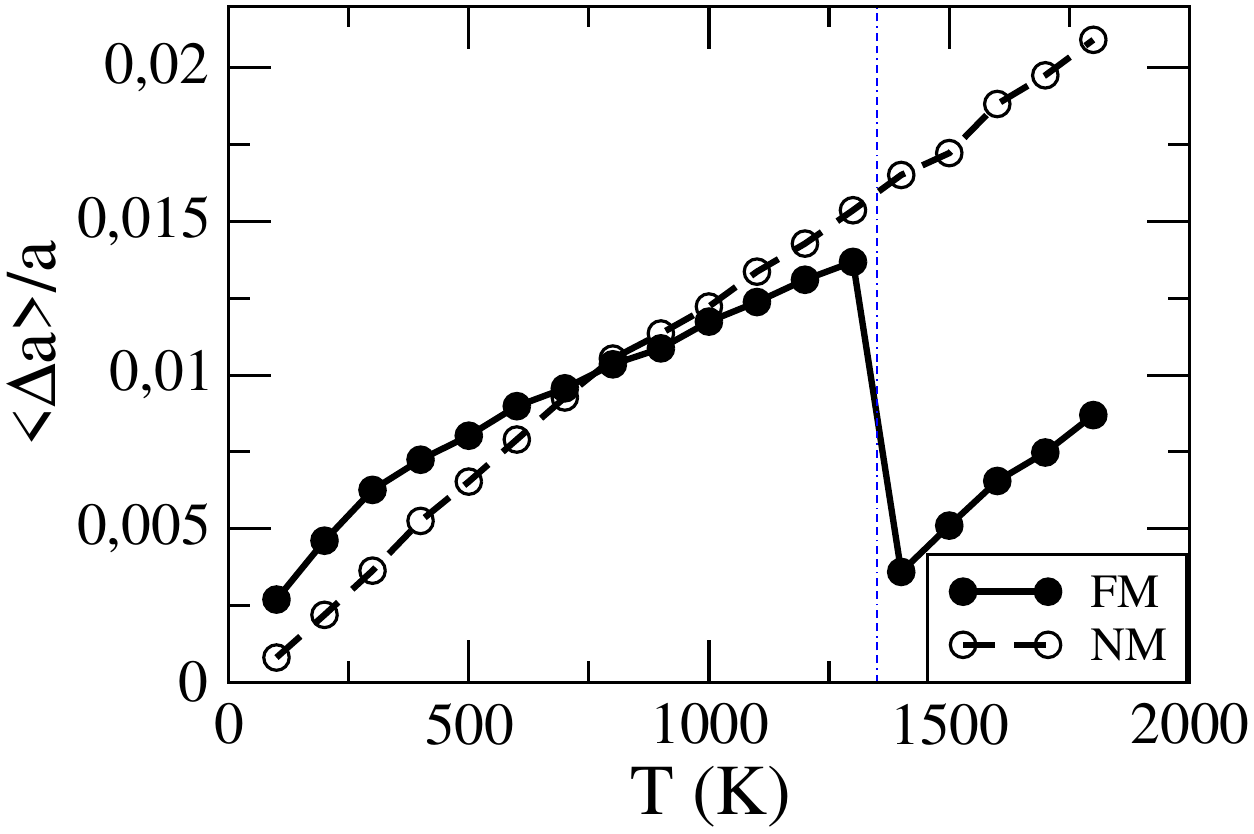}
\caption{Average linear thermal expansion coefficient for fcc Co as a function of temperature (FM and NM states). The red vertical line indicates the $\mathrm{T_{C}}$. }
\label{fgr:a_T}
\end{center}
\end{figure}
Therefore considering all degrees of freedom (magnetic, atomic and box relaxations), our TB model reproduces successfully the experimental Curie temperature emphasizing its remarkable ability to describe magnetic transition metals at finite temperature. As discussed above, the impact of the lattice relaxation is far from being negligible. To get an insight into this contribution, the linear thermal expansion of the fcc lattice $\left<\triangle a(T)\right>/a$ in FM and NM states is analysed and calculated as follows: 
\begin{equation}
\frac{\left<\triangle a(T)\right>}{a}= \frac{\left<a(T)\right>-a(T_{\mathrm{ref}})}{a(T_{\mathrm{ref}})}, 
\label{thermalexpansion}
\end{equation}
where $T_{\mathrm{ref}}=0$ K. In case of NM calculations, Fig.~\ref{fgr:a_T} illustrates the temperature dependence of the linear thermal expansion coefficient over a wide temperature range. A similar linear variation is observed for the FM states up to $\mathrm{T_{C}}$ where a contraction of the lattice parameter ($\sim$ 1\%) is found. This particular behavior is therefore a direct fingerprint of magnetism. According to our DFT and TB calculations at 0 K, the lattice parameters of the FM phases are always larger than in the non-magnetic ones (see Table \ref{tab:0K_E_I}). It is therefore tempting to think that the contraction observed at the Curie point is directly correlated to this difference in lattice parameters between both states. However, it is important to specify that after $\mathrm{T_{C}}$, the system tends towards a paramagnetic regime which is regarded as a collection of
disordered local moments. In this particular case, the random orientation of the spins results in the cancellation of the total magnetic moment. 
\begin{figure}[htbp!]
\begin{center}
\includegraphics[width=0.95\linewidth]{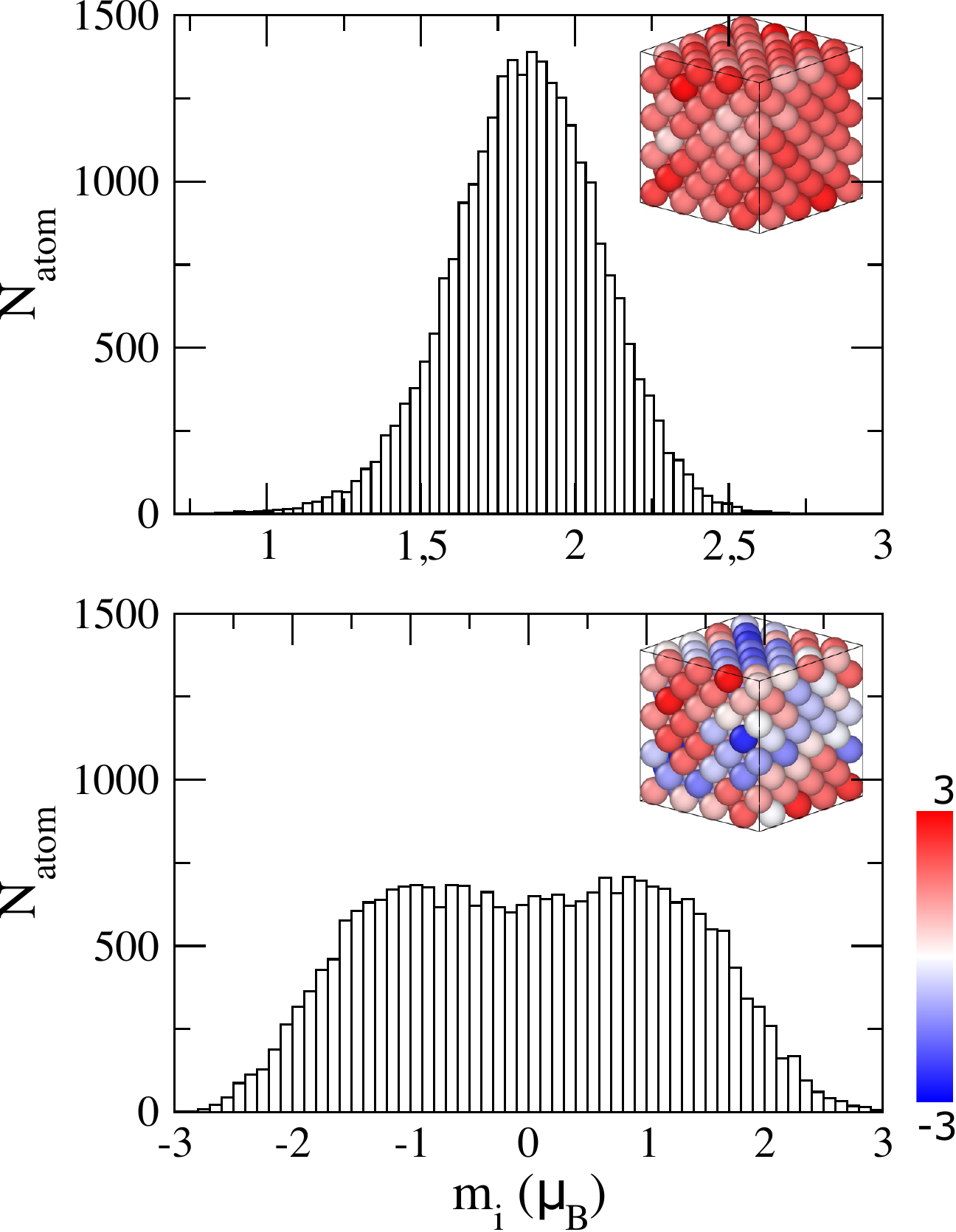}
\caption{Local magnetic moment distribution of 100 configurations of the fcc system below $\mathrm{T_{C}}$ at 1300 K (top) and above $\mathrm{T_{C}}$ at 1400 K (bottom) calculated by TB-FMA. One magnetic configuration of each temperature is represented with a color code scale from -3 $\mu_{B}$ to 3 $\mu_{B}$.}
\label{fgr:Tc_hist}
\end{center}
\end{figure}
Meanwhile, the NM state is characterized by the vanishing of all local moments. This  explains the different values obtained in the two calculations above $\mathrm{T_{C}}$ in Fig.~\ref{fgr:a_T}. To go deeper in this analysis, we consider the distribution of the local magnetic moments in the paramagnetic and ferromagnetic regimes. They are shown in Fig.~\ref{fgr:Tc_hist} for temperatures close to the phase transition: 50 K above and below the calculated Curie temperature. At 1300 K, a gaussian distribution centered around the value of the ground state magnetic moment ($m=1.87$ $\mu_{B}$), is observed with a dispersion due to thermal fluctuations, the typical signature of a ferromagnetic state. Above the Curie temperature, at 1400 K, the magnetic moment distribution is wider and its amplitude is lower, as expected in PM state. 
\begin{figure}[htbp!]
\begin{center}
\includegraphics[width=0.95\linewidth]{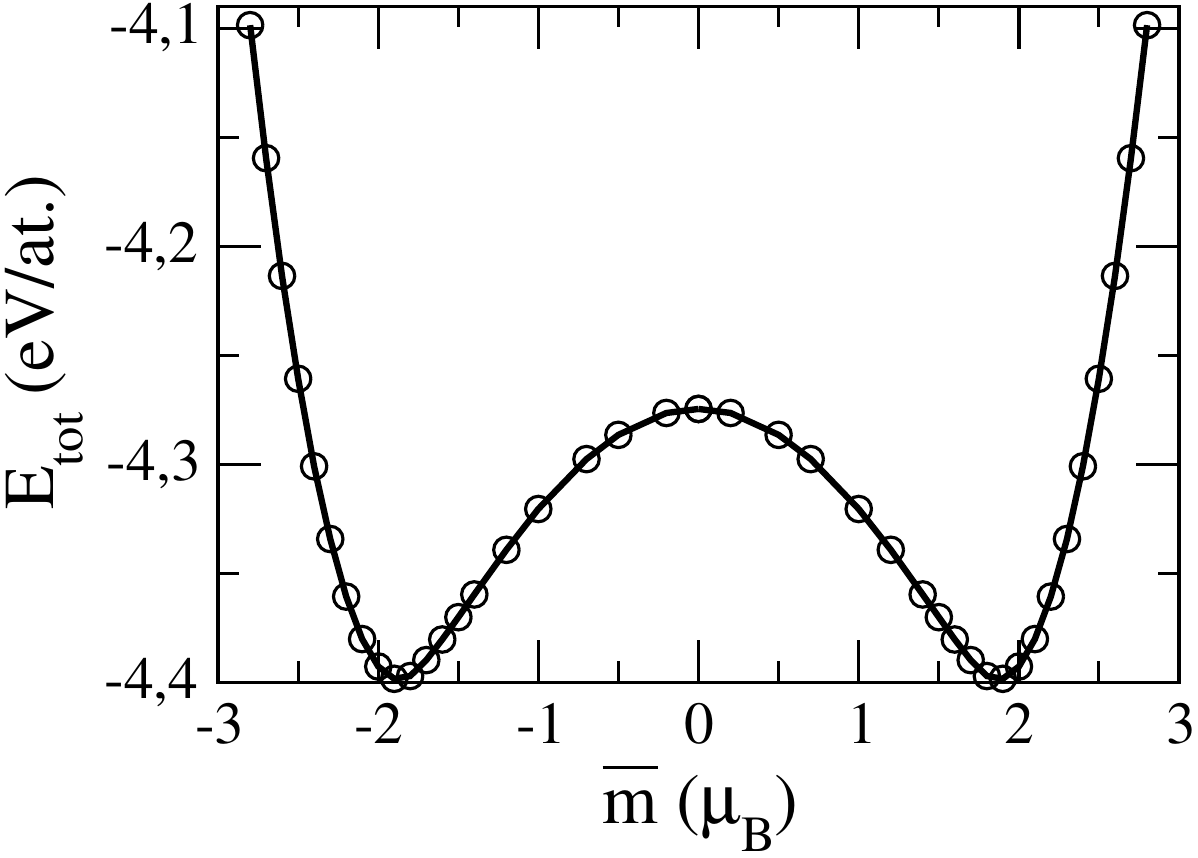}
\caption{Total energy as a function of the total magnetic moment for the fcc phase calculated by TB-FMA. The lattice parameter is kept constant and equal to its value at 0 K.}
\label{fgr:0K_E_m}
\end{center}
\end{figure}
To understand this behavior, it is useful to calculate the total energy (as defined in Eq.~\ref{Etotmag}) as a function of $m_{i}$~\cite{Ruban1} at 0 K. The results are presented in Fig.~\ref{fgr:0K_E_m}. The deep minimum at the positions of the on-site energies for the fcc phase results in a distribution of moments around this value below the Curie temperature. This explains the magnetic moment distribution of gaussian type centered on the minimum in the FM phase. We notice that the symmetrical distribution (centered around $m=-1.87$ $\mu_{B}$) may appear, depending on the MC moves. At higher temperature, the thermal excitations are large enough to avoid being trapped in the minimum. As a result, the magnetic moment is randomly distributed along the whole range of values leading to a more broadened shape centered on 0 $\mu_{B}$ as observed in Fig.~\ref{fgr:Tc_hist}. Interestingly, our TB-FMA interatomic potential coupled with specific MC trials on the magnetic moment turns out to be particularly successful in achieving randomly distributed collinear (up and down) magnetic moments. This is a direct result of our simulations and not an assumption established a priori, as in the disordered local moment (DLM) approximation.~\cite{Okatov2009, Razumovskiy2011} \\

Our study shows that atomic relaxations play an important role in the calculation of the T$_{c}$ of cobalt. Moreover, an abrupt variation of the lattice parameter is observed when the paramagnetic state is reached. In spite of such a great achievement, there is still an issue that seems to be problematic. In case of Co, it is well-known that the local on-site electron-electron interaction is strong enough to create local atomic moments fluctuating on a short-length scale.~\cite{Eich2017} Our TB-Stoner formalism behaves as an Ising state with a continuous distribution of moments contrary to the classical localized Heisenberg model. The latter is notoriously insufficient and is the source of much debate on localised versus itinerant magnetism, which can however be improved with effective interaction models (EIM) of a generalized-Heisenberg model. We will demonstrate that the two descriptions are not incompatible. Actually, such discrepancy might be due to the collinear approximation adopted in our formalism. Indeed, the high-temperature magnetic properties of cobalt are, at least in part, driven by fluctuating magnetic moments whose arrangement is intrinsically non-collinear. In our TB collinear approximation, only the amplitude of the magnetic moment fluctuates along one axis meaning that its three components are reduced to the longitudinal one. To justify such a choice, we will consider an effective interaction model on lattice based on a Heisenberg formalism with a non-collinear treatment. It will then be possible to decouple each contribution (longitudinal and transverse) of the magnetic moment and compare their distribution to the one from our purely collinear model. \\

The effective interaction model (EIM) is written as:
\begin{equation}
H = \frac{1}{2}\sum_{i}\sum_{j} V_{ij} + \frac{1}{2}\sum_{i}\sum_{j}J_{ij}\bold{m_{i}}\bold{m_{j}} \mbox{,}
\end{equation}
where $V_{ij}$ is a chemical pair interaction parameter between $i$-th and $j$-th atoms. The second term corresponds to the Heisenberg model where $J_{ij}$ is the exchange coupling parameter for the magnetic moment $\bold{m_{i}}$ and its neighbours $\bold{m_{j}}$. All these parameters are fitted to DFT calculations to investigate the phase stability in bcc Fe-Co systems.~\cite{Tranb} Consequently, the bcc phase of Co is considered in the following since there is no doubt that the general conclusions will be the same for fcc Co. More precisely, the second and the fifth neighbour have to be considered for the chemical interactions $V_{ij}$ and the magnetic interactions $J_{ij}$, respectively.~\cite{Tranb} In this model, the magnetic moment is described by a vector in the spherical coordinate system $\bold{m} = (m, \theta, \phi)$ with $m$ the amplitude, $\theta$ and $\phi$ respectively the polar and azimuthal angles. From DFT calculations~\cite{Tran2020} and experiments,~\cite{Collins} it has been shown that the average moment of Co atoms stays almost constant. Hence, the norm of the vector $\bold{m}$ is kept equal to 1.87 $\mu_{B}$ in coherence with our TB model. MC simulations are performed using the Metropolis algorithm with trials on the magnitude (Ising type $\pm m$), $\theta$ and $\phi$ angles. In these simulations, we perform $2.10^{3}$ macrosteps to let the system being equilibrated then we calculated the average of properties on $2.10^{3}$ macrosteps. We used a bcc system of 2000 atoms (10x10x10 cells) which is sufficiently large for the convergence of the energy and the magnetization.

\begin{figure}[htbp!]
\begin{center}
\includegraphics[width=1.00\linewidth]{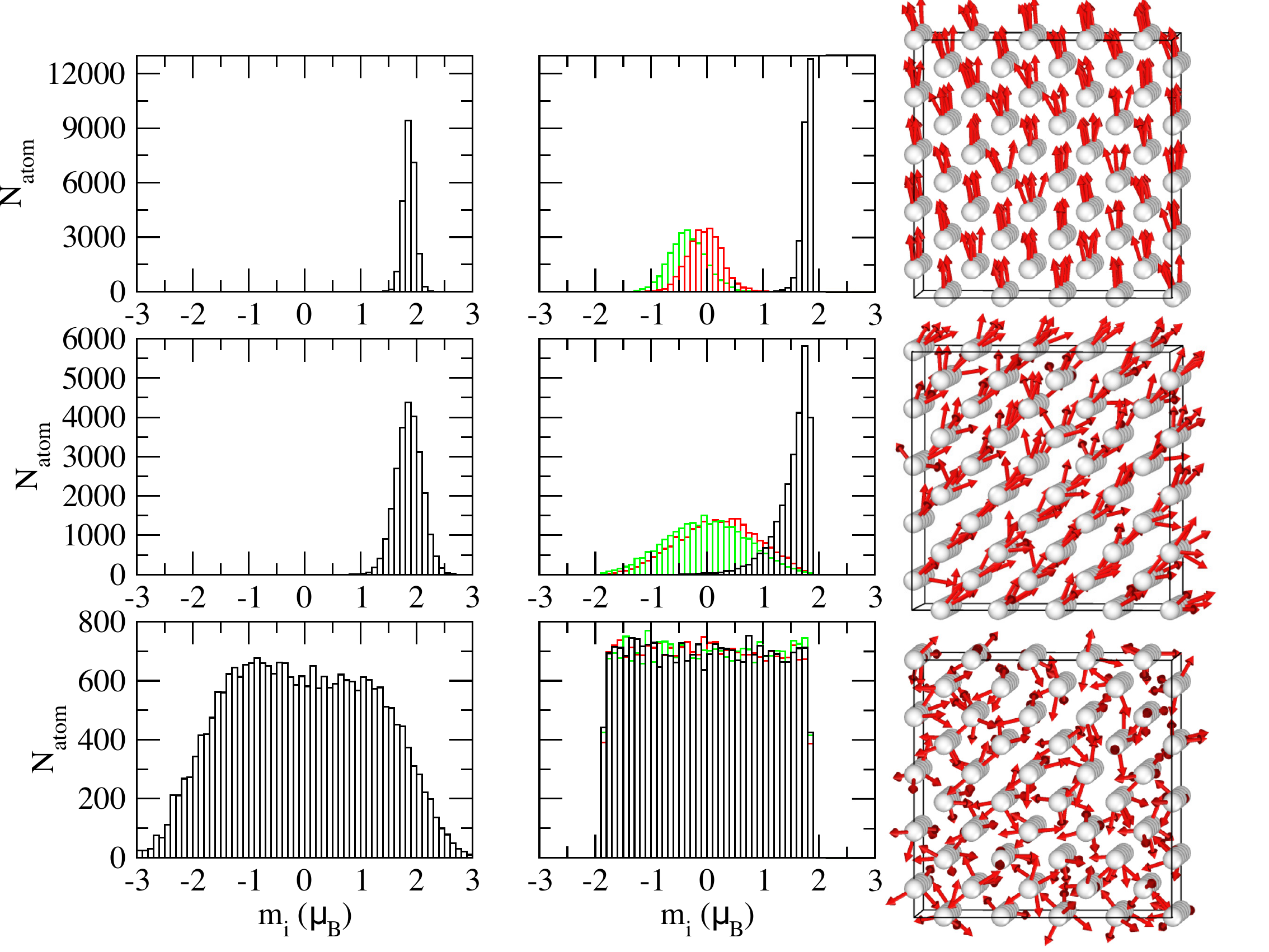}
\caption{Local magnetic moment distribution of 100 configurations of bcc system at 100 K (top), 500 K (middle) and 2000 K (bottom) from TB-FMA model (left column) and from EIM model (middle column) with the corresponding magnetic configuration (right column). In this EIM model, the magnetic moment is decomposed in three directions: $x$ in green, $y$ in red and $z$ in black.}
\label{fgr:comp_hist}
\end{center}
\end{figure}

In Fig.~\ref{fgr:comp_hist}, we report the magnetic distribution along the $x$, $y$ and $z$ axes. They are derived from the effective interaction model after projecting each component of the vector in the spherical coordinate system along the three orthogonal axes. Magnetic distribution at different temperatures are analysed and compared to the results from the TB-FMA model.
At low temperature, i.e. at 100 K, the magnetic moment distribution according to our TB model is characterized by a narrow gaussian distribution centered around 1.87 $\mu_{B}$. Regarding the non-collinear model, the $x$ and $y$ distributions are centered respectively on 0 $\mu_{B}$ and -0.5 $\mu_{B}$ leading to a slight shifting compared to the equilibrium value at 0 K whereas the $z$ distribution is well centered on 1.87 $\mu_{B}$. When increasing the temperature at 500 K, gaussian distributions become a little bit wider due to thermal fluctuations. On one, the magnetic distribution from TB-FMA is less impacted because the magnetic moment can only have longitudinal fluctuations. On the other, the distributions of the $x$ and $y$ coordinates are wider because of longitudinal and transverse spin fluctuations in the EIM model. Compared to the TB-FMA, $z$ distribution is similar except the maximal value of the magnetic moment which is limited to $\pm$1.87 $\mu_{B}$. Above the Curie temperature, the magnetic distribution is more dispersed than in the ferromagnetic state. Interestingly, the distribution resulting from the Heisenberg model is fully isotropic in good agreement with the collinear TB-FMA model. Only the tail of the gaussian is slightly different since in the non-collinear calculation it is quite sharp whereas it is much more spread out in the TB-FMA model. This difference comes from the approximation made in the Heisenberg model where the norm of the vector $\bold{m}$ is kept equal to 1.87 $\mu_{B}$ contrary to our TB approach. Thus our collinear spin approximation coupled with MC trials on both the atomic positions and the amplitude of the magnetism is in a more general way capable of capturing an important part of the magnetic excitation. The very complete study presented on the calculation of the Curie temperature thus fully validates the transferability of our interatomic potential as well as its ability to treat magnetic systems at finite temperature.

\section{A generalized-model for \textbf{magnetic} transition metals}
\label{Sec_Generalized_model}

A further benefit of a TB model is that it can be fairly easily generalized to other transition metal systems since we know qualitatively how the different parameters (transfer integrals, atomic energy levels) vary with the nature of the metallic element. In the following, we therefore take advantage of the physical transparency of the model to identify the influence of the Stoner parameter and of the number of electrons on the magnetic properties of transition metals in general. \\

\begin{figure}[htbp!]
\begin{center}
\includegraphics[width=1.0\linewidth]{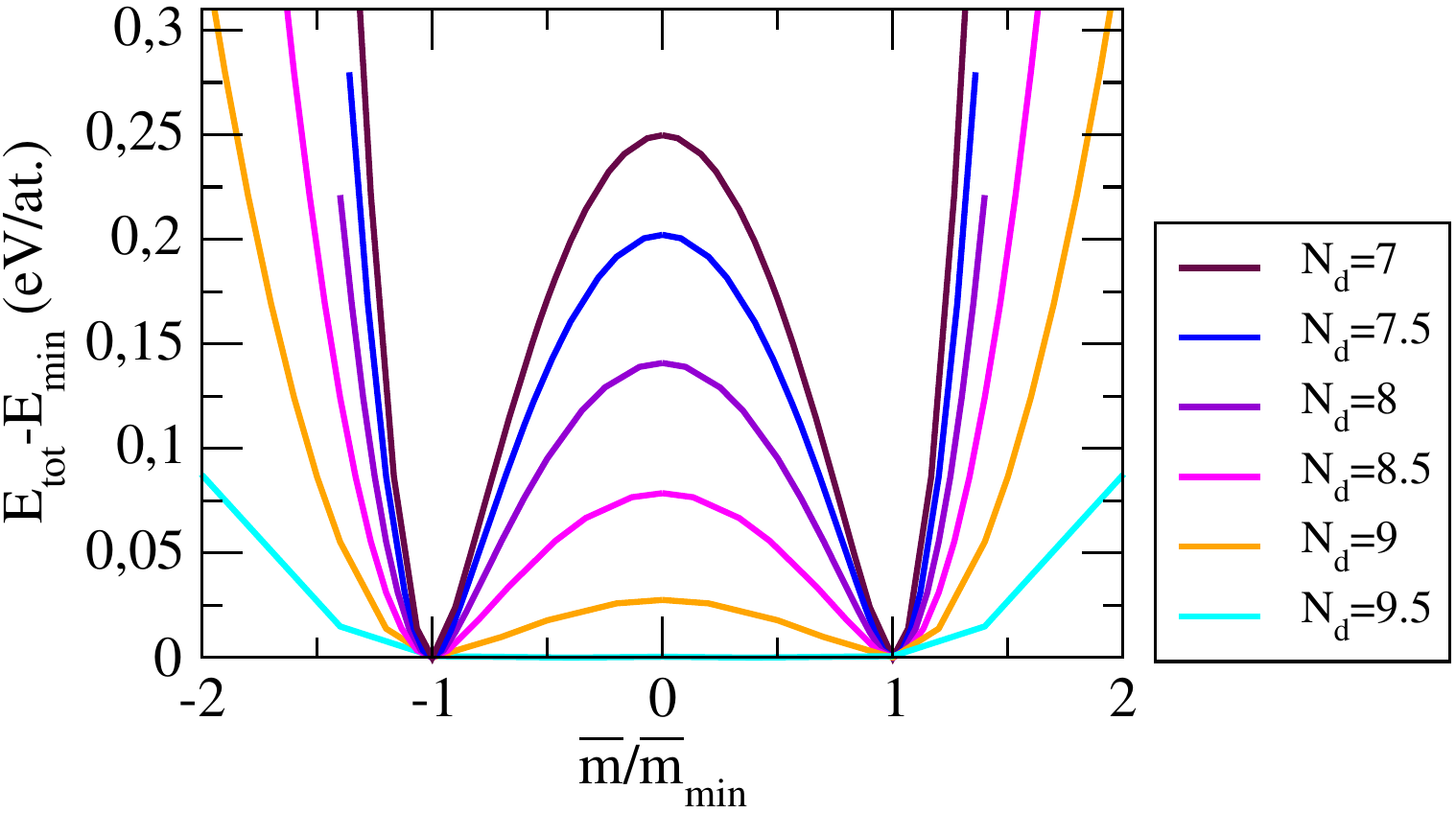}
\caption{Energy with respect to the ground state as a function of normalized magnetic moment for different $N_{d}$ at 0 K.}
\label{fgr:0K_E_Nd}
\end{center}
\end{figure}

First, a too small Stoner parameter results in a system in a non-magnetic state. Beyond a specific threshold for $I$, the Curie temperature increases linearly showing that T$_{C}$ can then be fitted to experimental measurements by simply tuning the Stoner parameter.
\begin{figure}[htbp!]
\begin{center}
\includegraphics[width=0.90\linewidth]{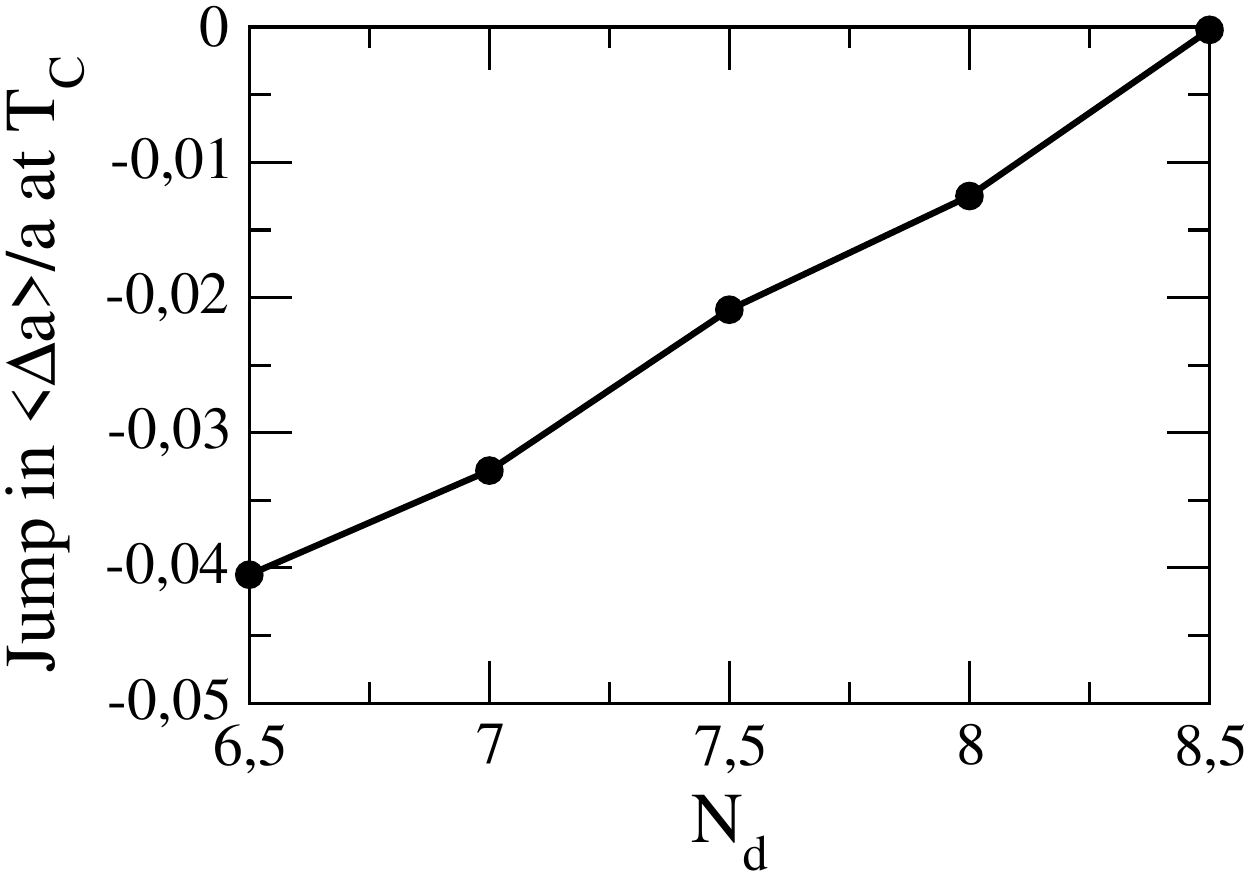}
\caption{Discontinuity of the thermal expansion coefficient at the Curie temperature as a function of the number of $d$ electrons.  $\Delta a$ is defined as $a_{\mathrm{PM}}-a_{\mathrm{FM}}$ respectively above and before the Curie point.}
\label{fgr:da_N}
\end{center}
\end{figure}
In a second step, we seek to investigate some specific magnetic properties in the ground state and at finite temperature as function of the number of $d$ valence electrons $N_{d}$. Obviously, a specific adjustment of all the parameters is mandatory to reproduce accurately the physical properties of the different transition-metal elements. However, since we are mainly interested in highlighting trends with band filling, all the parameters of the TB model are fixed.
\begin{figure}[htbp!]
\begin{center}
\includegraphics[width=1.0\linewidth]{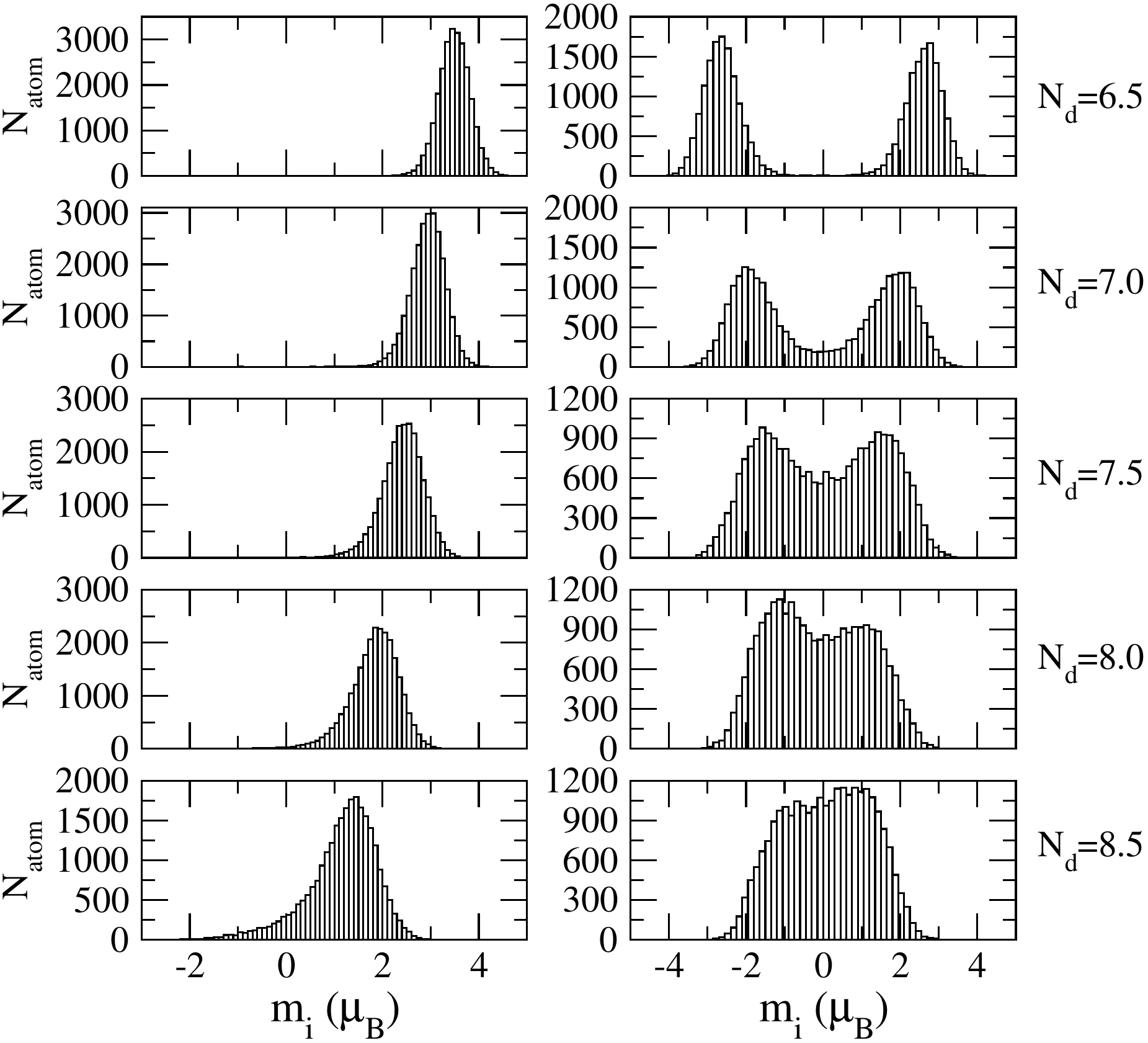}
\caption{Local magnetic moment distribution of 100 configurations of the fcc system below and above $\mathrm{T_{C}}$ for different $N_{\mathrm{d}}$.}
\label{fgr:T_N}
\end{center}
\end{figure}
In Fig. \ref{fgr:da_N}, the discontinuity of the linear thermal expansion at the Curie temperature for different $d$ band fillings is presented. By increasing the number of electrons, $\left<\Delta a\right>/a$ increases until $N_{\mathrm{d}}=8.5$ where it is almost zero. As discussed previously, this is related to the local magnetic moment distribution in Fig.~\ref{fgr:T_N} where a transition from a monomodal to a bimodal distribution is observed with the progressive filling of the $d$ band. 
Below the Curie temperature, the local magnetic moment distribution is always a gaussian centered on the magnetic moment whose value is determined by the number of electrons, explaining its shift towards lower values. One should keep in mind that the symmetrical distribution may appear depending on the MC moves. Above $\mathrm{T_{C}}$, two types of profile are identified in the PM state, i.e. a bimodal evolving towards a monomodal distribution with gradual filling of the $d$ states. At low $N_{\mathrm{d}}$, the local magnetic moment distributions are localized around plus or minus its mean absolute value with a dispersion due to thermal fluctuations. The sum is therefore zero corresponding to a PM state. When increasing $N_{\mathrm{d}}$, the bimodal type behavior tends to be reduced to a single and broad magnetic moment distribution centered around zero. Indeed, in Fig. \ref{fgr:0K_E_Nd}, the depth of minimum at the positions of the on-site energies decreases with gradual filling of the $d$ state. 
Therefore at high temperature, the thermal excitations are large enough to avoid being trapped in one local minimum. These results are in line with the widely accepted itinerant magnetism description of ferromagnetic transition metals.~\cite{Ruban1, Hubbard1, Hubbard2} In such case, a Heisenberg-type model is perfectly adapted to describe materials in which the electrons responsible for the magnetism are localized. Regarding  transition elements of the $3d$ series, this is typically the case for Fe or Co roughly corresponding to a band filling with $N_{\mathrm{d}}\leqslant7.0$ electrons. Besides, it is well-known that the theory of the Heisenberg ferromagnet fails entirely to describe magnetic properties of Ni. Regarding our analysis, this is in good agreement with the distribution of the longitudinal spin fluctuations which is no more localized when increasing $N_{\mathrm{d}}$. However, our results show how our TB-FMA combined with the Stoner formalism is adequat to describe the itinerant electron magnetism inherent to transition metals. Indeed, the local electronic structure description coupled to magnetic excitations driven by MC trials enables to unify a theory of itinerant electron magnetism at high temperatures. Different works have already been successful in establishing models to predict the magnetic properties at finite temperature of transition metals.~\cite{Moriya1978, Moriya1985, Ruban1}

Consequently, the variation of the lattice parameter at $\mathrm{T_{C}}$ appears to be a response of the system to the reorganization of local magnetic moments during the transition from FM to PM state. At small values of $N_{\mathrm{d}}$, the latter is very significant since it is characterized by the emergence of a second population of magnetic moments centered around a negative value of $m$. This strong transformation is associated with a very significant variation of the lattice parameter. For larger $N_{\mathrm{d}}$, the system goes smoothly from a rather sharp monomodal state to a wider monomodal distribution when crossing from FM to PM state. As a result, this  transformation does not involve a drastic variation of the lattice parameter. From an experimental point of view, a contraction of the lattice parameter at the Curie temperature has only been reported in case of Fe.~\cite{Touloukian1975} Anomalies at the Curie point characterized by an inflection of the slope have already been observed in pure elements~\cite{Ishida} or alloys such as Fe-Co~\cite{Stuart} as well as Fe$_{65}$Ni$_{35}$~\cite{Ruban}. The latter exhibits a very smooth temperature dependence with two different slopes below and above the Curie temperature, well known as the Invar effect which is a transition from a state with a higher magnetic moment and a large volume to a high temperature state with a lower magnetic moment and volume.

\section{Conclusion}

In this work we have presented a TB model based on the 4th moment approximation with an explicit magnetic contribution via the Stoner theory which provides an efficient tool to perform structural relaxations of magnetic transition metals. Remarkably, our approach coupled to MC simulations is able to reproduce localized and itinerant magnetism at finite temperature. The good agreement of our results for the Curie temperature in the case of Co highlights the importance of considering all physical contributions such as longitudinal spin fluctuations, atomic relaxations and lattice expansion. A further advantage of our model is that it can be fairly generalized to other magnetic transition metals since we know semiquantitatively how the different parameters (transfer integrals, atomic energy levels, Stoner parameter) vary with the nature of the metallic element. 

We have developed an interatomic potential which is efficient to investigate structural properties at finite temperature of large systems where complete relaxation is required. This is crucial in the case of nanoparticles where the Curie temperature decreases as the system size increases~\cite{Billas} because the magnetic moment is larger at the surface than in the core. Moreover, phase transformation at the nanoscale can also be driven by magnetic contribution and our model is able to capture this effect. In this context, the study of bcc-fcc transitions of Fe nanoparticles at different sizes is currently under investigation. An additional advantage of our model is that it can be extended to steel (Fe-C) and to transition metal alloys such as Co-Pt or Fe-Co where magnetism is a driving force of phase stability and chemical ordering. Lastly, a complete and precise description of the magnetism at a microscopic level can be developed by including non-collinear magnetism.~\cite{Ford2015} This work constitutes therefore a major step in the development of interatomic potential with a high degree of transferability to characterize phase transformation of magnetic transition metals. \\
 
\begin{acknowledgments}

This work was supported by the French-German ANR-DFG MAGIKID project. The authors thank ANR GiANT (N\degree ANR-18-CE09-0014-04). H.A and A.F thank Bernard Legrand for very helpful discussions. 

\end{acknowledgments}

\end{document}